\documentclass{aastex}
\usepackage{emulateapj5,psfig}

\newcommand{\etal}{{\it\ et\,al.\ }}
\newcommand{\beq}{\begin{equation}}
\newcommand{\eeq}{\end{equation}}
\newcommand{\maxima}{MAXIMA}

\newcommand{\microk}{$\mu{\mbox{K}}$}
\newcommand{\microkrtsec}{$\mu{\mbox{K}}\sqrt{\mbox{sec}}$}

\begin{document} 
\title{MAXIMA-1: A Measurement of the Cosmic
Microwave Background Anisotropy on angular scales of 
10\arcmin\ to 5\degr}

\author{S.~Hanany\altaffilmark{1,2}, P.~Ade\altaffilmark{3}, 
A.~Balbi\altaffilmark{4,2,15}, 
J.~Bock\altaffilmark{5,6}, J.~Borrill\altaffilmark{2,7},
A.~Boscaleri\altaffilmark{8}, P.~de Bernardis\altaffilmark{9}, 
P.~G.~Ferreira\altaffilmark{10,11}, V.~V.~Hristov\altaffilmark{6},
A.~H.~Jaffe\altaffilmark{2}, A.~E.~Lange\altaffilmark{2,6}, 
A.~T.~Lee\altaffilmark{14,15,2}, P.~D.~Mauskopf\altaffilmark{12},
C.~B.~Netterfield\altaffilmark{13}, S.~Oh\altaffilmark{2}, 
E.~Pascale\altaffilmark{8}, B.~Rabii\altaffilmark{2,14}, 
P.~L.~Richards\altaffilmark{2,14}, G.~F.~Smoot\altaffilmark{2,14,15,16},
R.~Stompor\altaffilmark{2,16,17}, C.~D.~Winant\altaffilmark{2,14}, 
J.~H.~P. ~Wu\altaffilmark{18} } 

\altaffiltext{1}{School of Physics and Astronomy, 
University of Minnesota/Twin Cities, Minneapolis, MN, USA}
\altaffiltext{2}{Center for Particle Astrophysics, University of
California, Berkeley, CA, USA}
\altaffiltext{3}{Queen Mary and Westfield College, London, UK}
\altaffiltext{4}{Dipartimento di Fisica, Universit\`a Tor Vergata, 
Roma, Italy} 
\altaffiltext{5}{Jet Propulsion Laboratory, Pasadena, CA, USA}
\altaffiltext{6}{California Institute of Technology, Pasadena, CA,
USA}
\altaffiltext{7}{National Energy Research
Scientific Computing Center, Lawrence Berkeley National Laboratory,
Berkeley, CA, USA}
\altaffiltext{8}{IROE-CNR, Firenze, Italy}
\altaffiltext{9}{Dipartimento di Fisica, Universit\`a La Sapienza,
Roma, Italy}
\altaffiltext{10}{Astrophysics, University of Oxford, Oxford, UK}
\altaffiltext{11}{CENTRA, Instituto Superior Tecnico, Lisboa, Portugal}
\altaffiltext{12}{Department of Physics and Astronomy, University of 
Wales, Cardiff, UK}
\altaffiltext{13}{Dept. of Physics and Astronomy, University of
Toronto, Canada}
\altaffiltext{14}{Dept. of Physics, University of California, Berkeley
CA, USA}
\altaffiltext{15}{Division of Physics,Lawrence Berkeley National Laboratory, 
Berkeley, CA, USA}
\altaffiltext{16}{Space Sciences Laboratory, University of Californina, 
Berkeley, CA, USA}
\altaffiltext{17}{Copernicus Atronomical Center, Warszawa, Poland}

\altaffiltext{18}{Department of Astronomy, University of California, 
Berkeley, CA, USA}

\begin{abstract}

We present a map and an angular power spectrum of the anisotropy of
the cosmic microwave background (CMB) from the first flight of MAXIMA.
MAXIMA is a balloon-borne experiment with an array of 16 bolometric
photometers operated at 100 mK.  MAXIMA observed a 124 deg$^2$ region
of the sky with 10\arcmin\ resolution at frequencies of 150, 240 and
410 GHz.  The data were calibrated using in-flight measurements of
the CMB dipole anisotropy.  
A map
of the CMB anisotropy was produced from three 150 and one 240 GHz
photometer without need for foreground subtractions.  Analysis of
this CMB map yields a power spectrum for the CMB anisotropy over the
range $36 \leq \ell \leq 785$.  The spectrum shows a peak with an
amplitude of $78 \pm 6$ \microk\ at $\ell \simeq 220$ and an amplitude
varying between $\sim 40$ \microk\ and $\sim 50$ \microk\ for 
$400 \la \ell \la 785$.

\end{abstract}


\section{Introduction} 

Measurements of the anisotropy of the cosmic microwave background
(CMB) can discriminate between cosmological models and determine
cosmological parameters with high accuracy \citep[and references
therein]{kamion_koso_99}.  Inflationary dark matter models, for
example, predict a series of peaks in the angular power spectrum of
the anisotropy.  The collected results from many experiments show
the existence of a first peak at angular scales corresponding
to the spherical harmonic multipole number $\ell \sim 200$. These
results have been interpreted as evidence for a flat universe
\citep{debernardis00,lange00,dodelson_knox00,tegmark_zald2000,pierpaoli_2000}. 
Additional observations probing a broad
range of angular scales would greatly increase confidence in these
results and further constrain cosmological parameters.

MAXIMA is a balloon-borne experiment optimized to map the CMB
anisotropy over hundreds of square degrees with an angular resolution
of 10\arcmin. In this paper we report results from the MAXIMA-1 flight
which took place on August 2, 1998.  These include a 100 square
degrees map of the CMB anisotropy and the resulting power spectrum
over the range $36 \leq \ell \leq 785$, which is the largest range
reported to date. Despite several common team members, the data
analysis was independent of that leading to the recently reported
BOOMERANG results \citep{debernardis00}.  A companion paper,
\citet{balbi_00}, discusses the cosmological significance of the
MAXIMA-1 results.

\section{Instrumentation}

\citet{lee_instrument} gives a detailed description of the MAXIMA
system.  It is based on a well-baffled, under-filled, off-axis
Gregorian telescope with a 1.3 m primary mirror, 
mounted on an attitude-controlled balloon-borne
platform.  A well-baffled
liquid-Helium-cooled optics box is lined with absorbing material
\citep{bock_goop} and contains two reimaging mirrors, low-pass
filters, field and aperture stops, feed horns for the 16 photometers,
and a focal-plane stimulator.  Eight conical single-mode horns
at 150 GHz and four multi-mode Winston horns each at 240 GHz and 410
GHz provide 10\arcmin\ beams at all three
frequencies. The frequency bands are defined with absorptive and
metal-mesh filters.  Radiation is detected with spider-web bolometers
\citep{bock_etal1996} operated at 0.1~K with an
adiabatic demagnetization refrigerator \citep{hagmann_95}.  
The bolometers are AC-biased to avoid low-frequency amplifier noise.
Additional channels with a constant
resistor, a thermometer, and a dark bolometer are used to monitor
electromagnetic interference, cross-talk, and drifts in electronic
gain and temperature.   
\begin{table*}
\begin{center}
\begin{tabular}{ccccc} \hline \hline
$\nu_{0}$ & $[\nu_{min},\nu_{max}]$ & FWHM$^{a}$ &$\tau$ & NET \\
   (GHz)  & (GHz)       & (arcmin)   &  (msec) & (\microkrtsec) \\ \hline
 150      & $[120,190]$ & (11.5,10)  & 10     & 80  \\ 
 150      & $[120,190]$ & (10.5,9.5)   & 7     & 90   \\ 
 150      & $[120,190]$ & (11.5,9.5) & 7      & 90  \\ 
 240      & $[190,270]$ & (12,8.5)   & 7      & 120  \\ 
 410      & $[380,430]$ & (11,8)     & 6      & 2050  \\ \hline 
\multicolumn{5}{l}{$^{a}$ The beams have some asymmetry. The FWHM} \\
\multicolumn{5}{l}{$\,\,\,$ are for the long and short axes. } \\ \hline
\end{tabular}
\end{center}
\caption{Central frequencies ($\nu_{0}$),
frequency bands ($\nu_{min}$,$\nu_{max}$),
beam full widths at half maximum (FWHM), detector time constants ($\tau$), 
and detector noise equivalent temperatures (NET) in CMB thermodynamic units
for the five \maxima-1 photometers discussed in this paper. The FWHM,
$\tau$, and NET were determined using flight data. The frequency bands
were measured in the laboratory.}  
\label{tab:experiment_parameters}
\end{table*}
The gondola azimuth is driven by a reaction wheel using information
from a two-axis magnetometer and a three-axis rate gyroscope.  The
telescope elevation is set using information from an angle encoder.
Observations were carried out at fixed elevation with the primary
mirror scanning $\pm \sim 2$\degr\ in azimuth at 0.45 Hz and 
the gondola also scanning in azimuth but at a frequency of $\sim 0.02$ Hz. 
Both scans were triangle functions of time with smoothed turnarounds. 

\section{Observations}

The MAXIMA-1 flight was launched from the National Scientific Balloon
Facility in Palestine, Texas at 1 UT on August 2, 1998.  Observations
of the CMB dipole for the purpose of calibration began at 3.6 UT when
the payload reached an altitude of 32 km and ended at 4.2 UT after
$\sim 100$ rotations at 3.3 rpm. The elevation angle was set to
51\degr. The payload reached float altitude of $38.4 \pm 0.4$ km at
4.6 UT.

The 1.6 hour CMB-1 observation began at 4.35 UT with a telescope
elevation of 46.3\degr.  The gondola was scanned $\pm 4.1$\degr\ in
azimuth at 16.1 mHz centered at 321.5\degr.  The 1.4 hour CMB-2
observation began at 6.0 UT with a telescope elevation of 32.3\degr.
The gondola was scanned $\pm 2.9$\degr\ at 21.3 mHz centered at
323\degr.  
Because of sky rotation, the combination of these
observations covered a nearly square region of the sky with an area of
124 square degrees of which 45\% is cross-linked at an angle of
$\sim$22\degr.

Observations of Jupiter were carried out from 7.5 to 8.1 UT to map the
telescope beams and provide additional calibration information.  The
elevation was fixed at 44.2\degr\ while sky rotation and the primary
mirror modulation provided $\sim 200$ transits across each beam.

\section{Pointing Reconstruction and Calibration}

We identified the stars which moved through the field of a CCD camera
aligned with the center of the primary mirror scan by using the balloon
location, telescope elevation, and the position of Polaris in an offset CCD
camera.  Interpolations using an angle encoder on the primary mirror,
rate gyroscopes, and the known star positions permitted pointing
reconstruction to better than 1\arcmin\ RMS. Less than 0.1\% of the
data had pointing uncertainty larger than 2\arcmin\ and were not used.

A full beam calibration of the 150 and 240 GHz photometers was
obtained from observations of the CMB dipole.  The data from each
rotation were $\chi^{2}$-fitted to a linear combination of a dipole model
\citep{lineweaver_96}, a galactic-dust emission model
\citep{schlegel_etal_98}, data from one 410 GHz photometer, an offset,
and a gradient.  The amplitude of each of these components was treated
as a free parameter.  A monotonic change in the detector calibration
of less than 9\% throughout the CMB observations, due to an increasing detector 
temperature, was monitored by illuminating the
focal plane with the stimulator lamp.  Estimated $1 \sigma$
calibration uncertainties were less than 4\% for each of the 150 and 240 
GHz photometers. The uncertainties in the dipole calibration and 
the time dependent calibration contributed
about equally to the total error, and systematic sources contributed
about 25\%. 

Beam maps and an independent calibration were obtained from
observations of Jupiter. We estimate a $1\sigma$ uncertainty of $\pm
0.5$\arcmin\ in the size of the beams. The beam profiles were
integrated and used with the angular diameter and brightness
temperature of Jupiter \citep{goldin_97} and the optical bandpass
functions to calibrate all 16 photometers.  For the data reported
here, the errors in the calibration from Jupiter were between 12 and
14\% (with about 10, 5, 5, and 2\% coming from uncertainties in the
beam solid angle, frequency bands, Jupiter's flux, and measured
signal, respectively).  The absolute calibrations from the dipole
agreed with those from Jupiter to within $1\sigma$, with the
Jupiter calibration predicting larger temperature fluctuations by 11 to
14\%.

\section{Map and Angular Power Spectrum}
\label{sec:map_and_power_spectra}

In this paper we report on the analysis of data from the three 150
GHz, one 240 GHz, and one 410 GHz photometer described in Table
\ref{tab:experiment_parameters}.  At 150 and 240 GHz these photometers
are the most sensitive in the \maxima\ array and give the highest
sensitivity of any CMB instrument reported to date.

The raw data for each photometer consisted of 2.3 million samples of
which about 16\% were not used. We removed the stimulator calibration
events and other events with an amplitude larger than $6\sigma$.
This procedure broke the data into 20 segments that were treated as
independent observations of the sky.  Samples in each of the segments
which were in excess of $4\sigma$, such as cosmic ray hits and short
telemetry drop-outs, were removed. For the data of one of the
photometers, we repeated the data analysis by using a threshold of
$3\sigma$ with no significant change in the resulting angular power
spectrum.  We deconvolved the transfer functions of the bolometers and
readout electronics and estimated the noise power spectrum from
sections of the time stream that had no gaps (Stompor et al., in
preparation).  We used the procedure of \citet{ferreira_jaffe_2000} to
confirm that the time-domain data are dominated by noise.  
We marginalized over frequencies lower than 0.1 Hz and higher than 
30 Hz, where we did not expect appreciable optical signals.

The calibrated time stream data were combined with the pointing
solution to produce a maximum likelihood pixelized map of temperature
anisotropy and a pixel-pixel noise correlation matrix for each
photometer \citep{wright_96,tegmark_97,bond_etal99}.  An area of $\sim
20$ square degrees of the map was not well cross-linked and is not
included in the present analysis.  The data showed a signal that was
phase-synchronous with the primary mirror scan which was due to
radio-frequency interference from on-board transmitters modulated by
the mirror motion.  This signal was constant within each data segment,
varied between different photometers, and had an amplitude of 100 -
300 \microk. We removed it by allocating fictitious map pixels to
values of the primary mirror angle and determined the maximum
likelihood map in these pixels simultaneously with the temperature
anisotropy map (Stompor et al., in preparation).

We verified that there are no noise correlations between maps of
different photometers by producing difference maps from the data of
pairs of photometers. The angular power spectra of these maps were
consistent with no signal.  We also showed that histograms of the
temperatures in the difference maps were consistent with the
distributions expected for no noise correlations at a
Kolmogorov-Smirnov significance level larger than 10\%.  A combined
temperature anisotropy map was then produced by adding individual maps
with a weight inversely proportional to their noise correlation
matrices. A Wiener filtered version of this map is shown in
Figure~\ref{fig:maxima1_map}.
We assign a calibration
uncertainty of 4\% to the magnitude of temperature fluctuations in the
combined map.

We calculated the angular power spectrum $C_{\ell}$ of the combined
map using the MADCAP \citep{borrill_1999} implementation of the
maximization of the likelihood function following \citet{bjk98}.  This
implementation assumes that the beam shape has axial symmetry.  We
produced an effective beam for the analysis of the combined map by
noise-weight averaging the individual beams. We followed the procedure
of \citet{wu_etal00} to find a symmetric approximation for the
effective beam and included the small smoothing provided by the
pixelization.  We tested this procedure for the \maxima-1 beams and
data and found no systematic bias of the $C_{\ell}$ estimates
\citep{wu_etal00} .  
We calculated the power spectrum of the temperature fluctuations using
15 bins in $\ell$ over the range $ 3 \la \ell \la 1500 $ assuming a
constant $\ell(\ell +1)C_{\ell}$ band power in each bin, and
marginalizing over the CMB monopole and dipole.  We further
marginalized over the bins at $\ell < 35$ and $\ell > 785$ and
diagonalized the $\ell$-bin correlation matrix using a variant of
techniques discussed in \citet{bjk98}.  The correlations between the
dominant bin and adjacent bins were typically less than 10\%.  Table
\ref{tab:cl_estimates} lists the dominant bins, the $C_{\ell}$ estimates, 
and the $\Delta T = \sqrt{\ell(\ell +1)C_{\ell}/2\pi}$ estimates for
the corresponding uncorrelated linear combinations of bins. We quote
$1 \sigma$ errors on the $C_{\ell}$ estimates assuming 68\% confidence
intervals using the offset log-normal distribution model of
\citet{bjk00}.  These errors do not include two additional independent
sources of uncertainty. 
Expressed as a percentage of $\ell(\ell+1)C_{\ell}/2\pi$, the
$1\sigma$ calibration error is a constant 8\% and the $\ell$-dependent
error due to the beam-diameter uncertainty, which has been shown to be 
fully correlated between $\ell$ bins, is given in
Table~\ref{tab:cl_estimates}. Information on the shape of the bin-power
likelihood functions and window functions will be made available on
the MAXIMA web site (http://cfpa.berkeley.edu/maxima).
\begin{table*}
\begin{center}
\begin{tabular}{cclcl} \hline \hline
$\ell_{eff}$ & $[\ell_{min},\ell_{max}]$ & $\ell(\ell+1)C_{\ell} /2\pi$ & Beam Error & $\Delta T$ \\
   &         & ($\mu K^{2}$) & (\%) & (\microk) \\ \hline 
 77 & [  36, 110]& $2000_{-510}^{+680} $ & $\pm 0$ & $45_{-6}^{+7} $ \\
147 & [ 111, 185]& $2960_{-550}^{+680} $ & $\pm 0.5$ & $54_{-5}^{+6} $ \\
223 & [ 186, 260]& $6070_{-900}^{+1040}$ & $\pm 1.5$ & $78_{-6}^{+6} $ \\
300 & [ 261, 335]& $3720_{-540}^{+620} $ & $\pm 2.5$ & $61_{-5}^{+5} $ \\
374 & [ 336, 410]& $2270_{-340}^{+390} $ & $\pm 3.5$ & $48_{-4}^{+4} $ \\
447 & [ 411, 485]& $1530_{-270}^{+310} $ & $^{+5}_{-4.5}$ & $39_{-4}^{+4} $ \\
522 & [ 486, 560]& $2340_{-380}^{+430} $ & $^{+6.5}_{-6}$ & $48_{-4}^{+4} $ \\
597 & [ 561, 635]& $1530_{-340}^{+380} $ & $^{+8}_{-7} $  & $39_{-5}^{+5} $ \\
671 & [ 636, 710]& $1830_{-440}^{+490} $ & $^{+9.5}_{-8.5}$ & $43_{-5}^{+5} $ \\
742 & [ 711, 785]& $2180_{-620}^{+700} $ & $^{+11}_{-10}$ & $47_{-7}^{+7} $ \\
\hline
\end{tabular}
\end{center}
\caption{
	Uncorrelated Power Spectrum from the MAXIMA-1 map. Errors
	are the 68\% integrated probability of the offset
	log-normal likelihood functions with a constant prior in
	either $\ell(\ell+1)C_{\ell}/2\pi$ or $\Delta
	T=\sqrt{\ell(\ell +1)C_{\ell}/2\pi} $ for their respective
	columns.  Here $[\ell_{min},\ell_{max}]$ gives  
	the ranges of the dominant bins. The correlated
	beam errors are described in the text.   
}
\label{tab:cl_estimates}
\end{table*}

The top panel of Figure \ref{fig:power_spectrum} shows the maximum
likelihood power spectrum, an inflationary adiabatic model that best
fits the MAXIMA-1 and COBE/DMR power spectra, and a $\Lambda$CDM
model. 
The $\chi^{2}$
for the best fit and for the $\Lambda$CDM models are 41 and 50,
respectively, using all 38 data points. If we only use the 10 data
points of MAXIMA-1 we obtain $\chi^2 = 7$ and 9 for the best fit 
and  $\Lambda$CDM, respectively \citep{balbi_00}.

\section {Foregrounds}

Foreground sources include emission from the earth, the atmosphere,
interstellar dust, free-free radiation, synchrotron radiation, and
point sources, and scattering due to the Sunyaev-Zeldovich effect.
The two CMB scans, performed with a separation of 1.5 hours at
different telescope elevations, show consistent structure. This
temporal stability is inconsistent with an atmospheric or ground-based
origin for the signal.  We extrapolated the 100 \micron, 10\arcmin\
resolution \citet{schlegel_etal_98} map of our observed region to
lower frequencies using the two-component dust model favored by
\citet{fink_1999}. The predicted RMS dust temperature is only 2.3 and 9.3
\microk, for the 150 and 240 GHz bands, respectively. We found a
statistically significant correlation between the dust model and our
maps.  The ratio of the detected to predicted RMS
dust signal was statistically consistent with unity.
When we subtracted the correlated dust signal from the maps the change
in the measured angular power spectrum was negligible.  A catalog
search \citep{sokasian_etal2000,gawiser_smoot1997} yielded no
detectable radio or infra-red sources in any of the frequency bands.
Estimates of bremsstrahlung and synchrotron radiation
\citep{bouchet_gispert_1999} yielded contributions of less than 1
\microk\ at 150 and 240 GHz and no subtractions were made.  

Integrating the measured power spectrum of one photometer at 150 GHz
and the one at 240 GHz we find for a thermodynamic temperature
fluctuations ratio of $0.91 \pm 0.18$ compared to unity for the CMB,
0.06 for emission from dust, 2.1 for synchrotron, and 3.7 for
free-free radiation, assuming the \citet{tegmark_foreg} ``middle''
foregrounds model.  For the ratio between 150 and 410 GHz, we found a
lower limit of $1.1 \cdot 10^{-2}$ compared to $3.2 \cdot 10^{-4}$ for
dust and $5.7 \cdot 10^{-5}$ for atmosphere. The observed anisotropy
is not consistent with the thermal SZ effect, which would produce
anti-correlated structure in the 150 and 240 GHz bands.

\section{Tests for Systematic Errors}

Because of computational limitations, tests for systematic errors were
done with maps which had square pixels of 8\arcmin\ and 10\arcmin\ on
a side. The power spectra calculated from the 8\arcmin-pixel and
5\arcmin-pixel combined maps were statistically consistent.  The
following combinations of the data were analyzed and produced a power
spectrum consistent with no signal: (1) a dark bolometer, (2) the data
from the 410 GHz photometer, (3) the difference between the
overlapping part of the combined map from the CMB-1 and -2 scans, (4)
the differences between the maps produced by different photometers.
We also weight-averaged the maps of the second and third photometers
in Table~\ref{tab:experiment_parameters} and the first and fourth. The
maximum likelihood estimate of the power spectrum of the difference
between these independent maps is consistent with no signal as shown
by the open circles in the top panel of Figure~\ref{fig:power_spectrum}.

We compared the estimate of the angular power spectrum to that
obtained using: (1) only the sections of the map where the CMB-1 and
-2 scans overlap, (2) a map of each of the CMB scans alone.  We made
maps and calculated $C_{\ell}$ estimates from the data of each
photometer alone and using: (1) only a sub-section of the time stream
data, (2) a high-pass filtered version of the time stream where the
high-pass was a time domain box-car with a width of 10 sec, (3)
various combinations of frequency marginalizations between 30 and 70
Hz, and 0.05 and 0.3 Hz, respectively. In all these cases the computed
power spectra agreed among themselves and with the power spectrum
presented in this paper.  The Kolmogorov-Smirnov test (see Section
\ref{sec:map_and_power_spectra}), as applied to difference maps and to
a map of the dark bolometer, confirms that we correctly estimated the
noise in the experiment and that the pixel-domain noise is Gaussian.

We used simulations to test the algorithm used to subtract the
signal that was phase-synchronous with the primary mirror
modulation. We found that the power spectrum estimate was not biased
with phase-synchronous signals larger than those observed in the data.
If we make the maps without removing the phase-synchronous signal the
power spectrum estimate changes only at $\ell \la 200$.

The computer programs used to generate the maps and power spectra were
tested extensively using simulations of the time domain data and
noise. Maps were produced by two independent computer codes and the
power spectra calculated from these maps were consistent.  We used one
of the map-making codes to make maps of Jupiter and found them
consistent with those obtained with a simple data-binning technique.

\section{Discussion}

We have observed temperature anisotropy on the sky at 150 and 240 GHz
that is consistent with fluctuations in the cosmic microwave
background radiation, and inconsistent with any known foreground.  The
observations were carried out with photometers that give the highest
CMB sensitivity reported to date. Our measurements cover a range of
angular scales corresponding to the multipole range $36 \leq \ell \leq
785$, which is the largest yet reported by a single experiment.  The
measured angular power spectrum shows a clear peak at $\ell \simeq
220$, and an amplitude varying between $\sim 40$ \microk\ and $\sim
50$ \microk\ at $400 \la \ell \la 785$.  The power spectrum is well
fit by an inflationary adiabatic model over the entire range of
$\ell$. The best-fit model has a total energy density close to unity
and a non-zero cosmological constant.  The \maxima-1 power spectrum
appears consistent with that of the BOOMERANG experiment
\citep{debernardis00} once the power spectra of the two experiments
are scaled by factors equal to their respective $1 \sigma$ calibration
uncertainties, see Figure~\ref{fig:power_spectrum}.  
A detailed analysis of the combined data sets which
includes a determination of the calibration factors that bring
the experiments to agreement is presented in \citet{jaffe_maxiboom}.

\acknowledgments

We thank Danny Ball and the other staff at NASA's National Scientific
Balloon Facility in Palestine, TX for their outstanding support of the
MAXIMA program.  MAXIMA is supported by NASA Grants NAG5-3941,
NAG5-6552, NAG5-4454, GSRP-031, and GSRP-032, and by the NSF through
the Center for Particle Astrophysics at UC Berkeley, NSF cooperative
agreement AST-9120005, and a KDI grant 9872979.  The data analysis
used resources of the National Energy Research Scientific Computing
center which is supported by the Office of Science of the
U.S. Department of Energy under contract no.\ DE-AC03-76SF00098. PA
acknowledges support from PPARC rolling grant, UK.

\clearpage

\setcounter{figure}{0}

\begin{figure}
\plotone{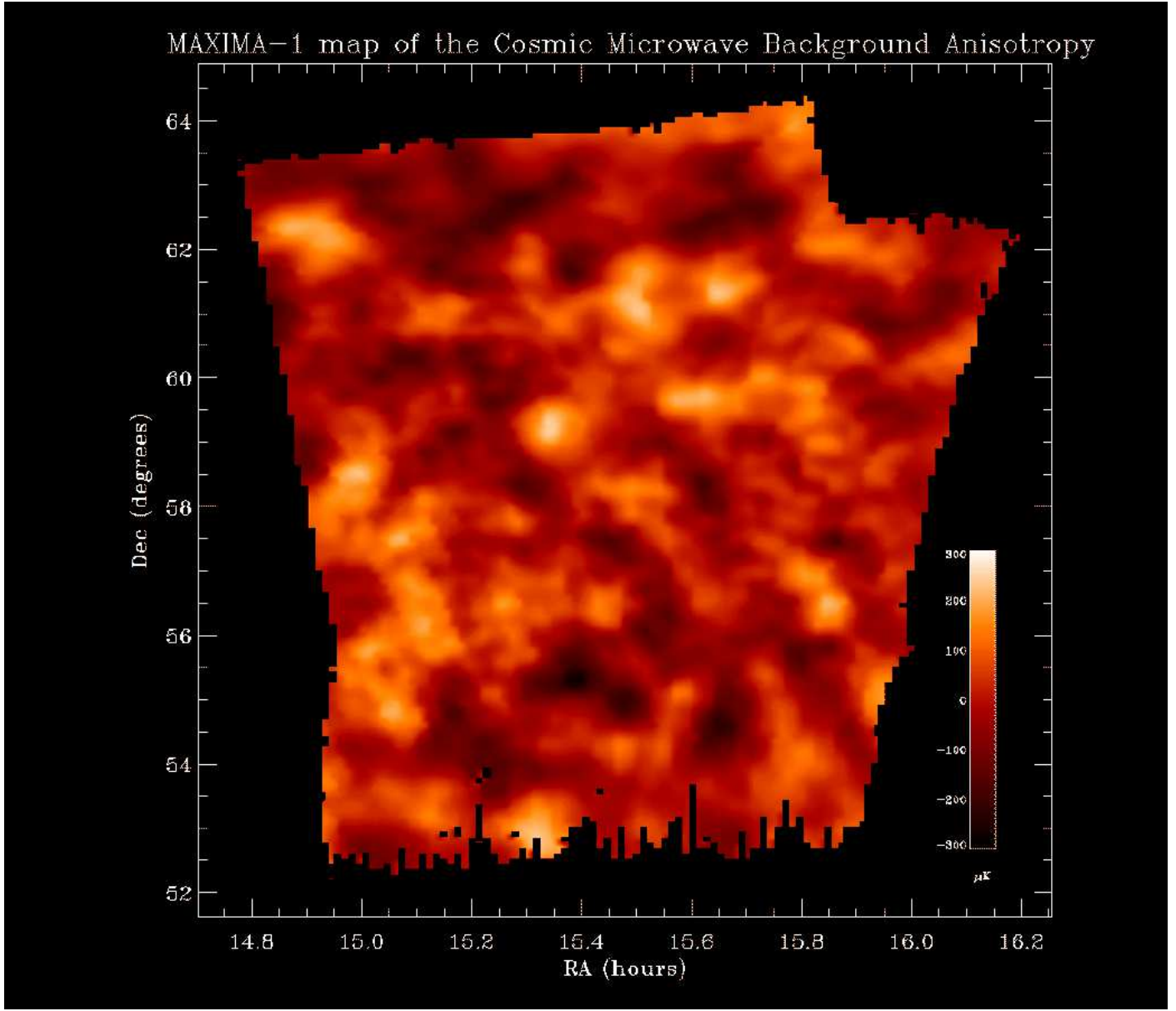}
\caption{A Wiener filtered map of the CMB anisotropy with 
10\arcmin\ resolution from MAXIMA-1.
The map is made using data from three 150 GHz and one 240 GHz
photometer and contains 15,000 5\arcmin$\times$5\arcmin\ pixels.  
We used the angular power spectrum shown
in the top panel of Figure~\ref{fig:power_spectrum} as the prior 
for the Wiener filter. 
Pixel boundaries have been smoothed using interpolation. }
\label{fig:maxima1_map}
\end{figure}

\begin{figure}
\centerline{\psfig{file=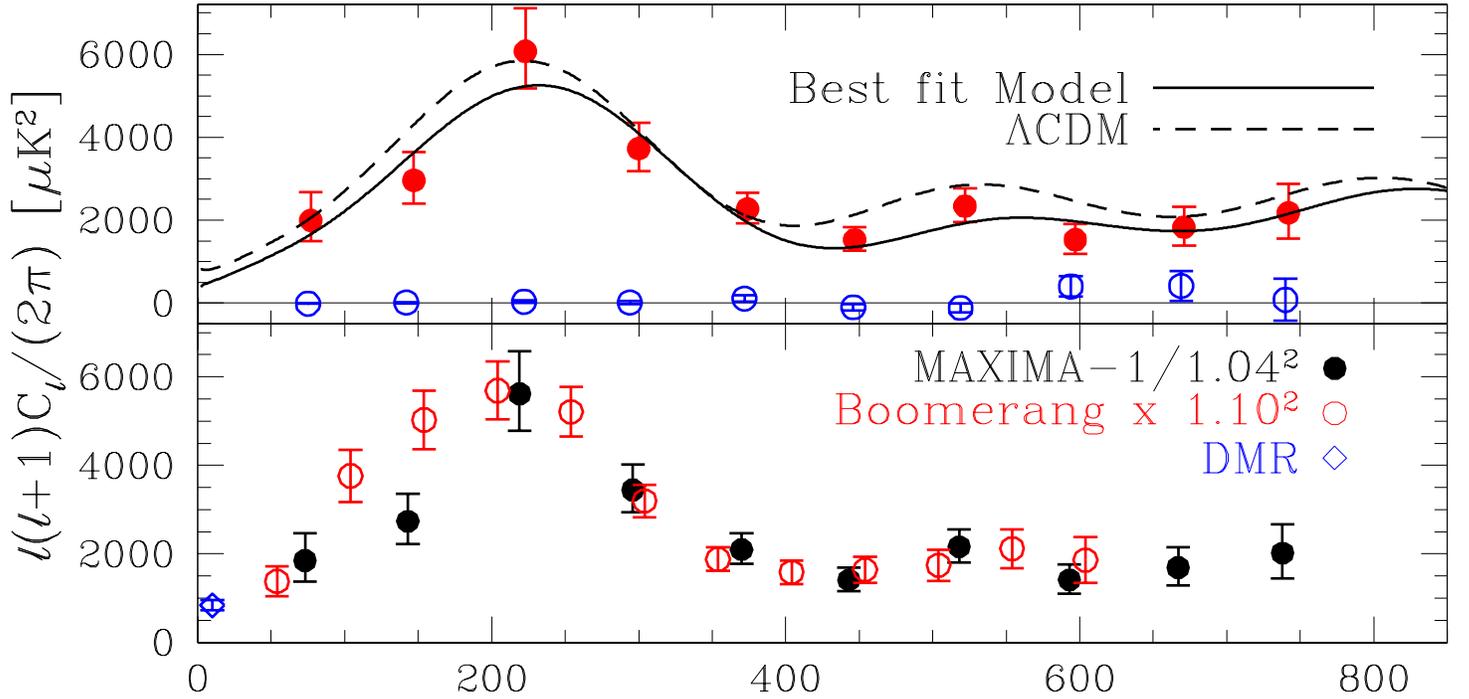} }
\caption{Top panel: 
Angular power spectrum of the CMB anisotropy from the MAXIMA-1 map
shown, in Figure~\ref{fig:maxima1_map} (filled circles).  The error
bars are 68\% confidence intervals calculated using the offset
log-normal likelihood functions of \citet{bjk00}.  The solid (dashed)
curve is the best fit ($\Lambda$CDM) inflationary adiabatic cosmology
to the MAXIMA-1 and COBE/DMR data. The models have ($\Omega_{b}$,
$\Omega_{cdm}$, $\Omega_{\Lambda}$, $n$, $h$) = (0.1, 0.6, 0.3, 1.08,
0.53), and (0.05, 0.35, 0.6,1 ,0.65), respectively \citep{balbi_00}.
The open circles are the estimated power spectrum of the difference
between two independent maps, each produced by a weighted-average of
the maps from a pair of photometers.  Bottom panel: A comparison of
the MAXIMA power spectrum with that of the recently reported BOOMERANG
experiment \citep{debernardis00}.  Consistency between the power
spectra has been achieved by scaling the MAXIMA-1 (BOOMERANG) 
power spectrum down (up)
by a factor equal to its 8\% (20\%) $1\sigma$ calibration uncertainty.
These data show a suggestion of a peak at
$\ell \sim 525$. 
}
\label{fig:power_spectrum}
\end{figure}

\end{document}